\documentclass{aa}  

\usepackage{graphicx}
\usepackage{natbib}
\usepackage{threeparttable}
\usepackage{txfonts}
\begin{document}

 \newcommand{\betacrbbold}{\mbox{\boldmath $\beta$}~CrB} 

 \newcommand{\betacrb}{$\beta$~CrB}
 \newcommand{\tenaql}{10~Aql}
 \newcommand{\gammaequ}{$\gamma$~Equ}

 \newcommand{\thetald}{$\theta_{\rm LD}$}
 \newcommand{\fbol}{$f_{\rm bol}$}
 \newcommand{\acir}{$\alpha$~Cir}
 \newcommand{\ie}{i.e.}
 \newcommand{\eg}{e.g.}
 \newcommand{\cf}{cf.}
 \newcommand{\kms}{km\,s$^{-1}$}
 \newcommand{\teff}{\ensuremath{T_{\rm eff}}} 
 \newcommand{\teffii}{T_\rm{eff}}
 \newcommand{\teffsun}{$T_\rm{eff},{\odot}$}
 \newcommand{\logg}{\ensuremath{\log g}}
 \newcommand{\feh}{[Fe/H]}
 \newcommand{\sunsym}{$\odot$}
 \newcommand{\msun}{${\rm M}_{\odot}$}
 \newcommand{\rsun}{${\rm R}_{\odot}$}
 \newcommand{\lsun}{${\rm L}_{\odot}$}
 \newcommand{\percent}{\,{\%}}
 \newcommand{\kepler}{{\em Kepler}}
 \newcommand{\corot}{{\em CoRoT}}
 \def\note #1]{{\bf #1]}}
 \def\muHz{\, \mu{\rm Hz}}

\renewcommand{\sun}{\odot}
\newcommand{\Teffsun}{{T_{\mathrm{eff},\sun}}}
\newcommand{\Teff}{{T_{\mathrm{eff}}}}

 \newcommand{\vsini}{\ensuremath{v \sin i}}
 \newcommand{\feone}{Fe\,{\sc I}}
 \newcommand{\fetwo}{Fe\,{\sc II}}
 \newcommand{\kic}{KIC\,8410637}

\title{The mass and age of the first SONG target: the red giant 46 LMi.
\thanks{Based on observations made with the Hertzsprung SONG telescope operated at
the Spanish Observatorio del Teide on the island of Tenerife by the Aarhus and Copenhagen Universities
and by the Instituto de Astrof\'{\i}sica de Canarias.}}
\titlerunning{First SONG target: 46 LMi}
\authorrunning{S. Frandsen et al.}

\author{
S.~Frandsen\inst{1}
\and 
M.~Fredslund Andersen\inst{1}
\and
K.~Brogaard\inst{1}
\and
C.~Jiang\inst{2}
\and
T.~Arentoft\inst{1}
\and
F.~Grundahl\inst{1}
\and
H.~Kjeldsen\inst{1}
\and
J.~Christensen-Dalsgaard\inst{1}
\and
E.~Weiss\inst{1}
\and
P.~Pall\'e\inst{3}
\and
V.~Antoci\inst{1}
\and
P.~Kj{\ae}rgaard\inst{4}
\and
A.~N.~S{\o}rensen\inst{4}
\and
J.~Skottfelt\inst{6,5}
\and U.~G.~J{\o}rgensen\inst{5}
} 
%
\institute{
Stellar Astrophysics Centre, Department of Physics and Astronomy, Aarhus University, 
Ny Munkegade 120, DK-8000 Aarhus C, Denmark  \and
School of Physics and Astronomy, Sun Yat-sen University, 2 Daxue Road, Tangjia, Zhuhai 519082, Guangdong Province, China \and 
Instituto de Astrof\'isica de Canarias (IAC), 38200 La Laguna, Tenerife, Spain \and 
Niels Bohr Institute, University of Copenhagen, Juliane Maries Vej 30, 2100 
K{\o}benhavn {\O}, Denmark \and 
Niels Bohr Institute \& Centre for Star and Planet Formation, University of Copenhagen, 
{\O}ster Voldgade 5, DK-1350 -- Copenhagen K, Denmark \and
Centre for Electronic Imaging, Dept. of Physical Sciences. The Open
University, Milton Keynes MK7 6AA, UK
}

\date{Received ; Accepted }
 
\abstract
{
The Stellar Observation Network Group (SONG) is an initiative to build
a worldwide network of 1m telescopes with high-precision radial-velocity
spectrographs.
Here we analyse the first radial-velocity time series of a red-giant  star measured by the SONG telescope at Tenerife. 
The asteroseismic results demonstrate a major increase in the achievable precision of the parameters for red-giant stars  obtainable from ground-based observations.
Reliable tests of the validity of these results are  needed, however,  before the accuracy of the parameters can be trusted.
}
{We analyse the first SONG time series 
for the star 46~LMi, which has a precise parallax and an angular diameter measured from interferometry, and therefore a good determination of the stellar radius.
We use asteroseismic scaling relations to obtain an accurate mass, and modelling to determine the age.
}
{
A 55-day time series of high-resolution, high S/N spectra were obtained
with the first SONG telescope. We derive the asteroseismic parameters by analysing the power spectrum. To give a best guess on the large separation
of modes in the power spectrum, we have applied a new method which uses the scaling of \kepler\ red-giant stars to 46~LMi.
}
{Several methods have been
applied: classical estimates, seismic methods using the observed time series, and
model calculations to derive the fundamental parameters of 46~LMi. 
Parameters determined using the different methods are consistent within the uncertainties.
We find the following values for the mass $M$ (scaling), radius $R$ (classical), age (modelling), and surface gravity (combining mass and radius): $M = 1.09\pm0.04$\,\msun , 
$R = 7.95\pm0.11$\,\rsun\, age $t = 8.2\pm1.9$\,Gy, and $\log g = 2.674 \pm 0.013$.
}
{
The exciting possibilities for ground-based asteroseismology of solar-like
oscillations with a fully robotic network have been illustrated with the
results obtained from just a single site of the SONG network. 
The window function is still a severe problem which will be solved when there are more nodes in the network.
}


   \maketitle
%

\section{Introduction \label{sec:intro}}

Asteroseismology has made a big step forward thanks to space-borne 
photometry missions. 
This  has been a revolution, particularly for red-giant  stars (\citealt{bedding1}; \citealt{beck}; \citealt{joergen}), as the need for very long time series 
has been solved by the long-duration observations of \corot\ and \kepler .
Ground-based campaigns suffer from much shorter time coverage, and even for 
long campaigns \citep{procyon} we are far from the time coverage of 
space-borne missions. Furthermore, campaigns must  use high-precision radial
velocities to detect the stellar oscillations since
ground-based photometry is unable to obtain the required 
precision \citep{stello2006,stello2007}.

However, it is still useful to conduct ground-based radial-velocity campaigns, 
if we can use dedicated networks of telescopes which can give close to 24-hour 
coverage for extended periods as discussed in detail by \cite{multiple}.
By measuring stellar oscillations using 
radial velocities the effects of surface granulation are strongly reduced compared
to photometric measurements. This allows us to detect modes at
lower frequencies, and modes of $l\,=\,3$ are easier to detect using 
radial velocities. 

We  are particularly interested in nearby stars where an angular diameter can be
measured accurately with an interferometer.
The idea of combining interferometric measurements with observations
of solar-type oscillations in the context of red-giant stars was employed for the star $\epsilon$~Oph (\citealt{mazum2009}),
where the asteroseismic data were obtained from space with the photometric MOST mission \citep{most}.
This was a favourable case because the angular diameter was
measured with high accuracy and the space data do not suffer from
atmospheric and alias problems. This is reflected in the good agreement
found between the interferometric radius and the asteroseismic radius.
Another similar study has been presented by \cite{beck2015} for the red-giant stars
$\gamma$~Per and $\theta^1$~Tau, where the asteroseismic
measurements come from multisite radial-velocity measurements.

The Stellar Observations Network Group (SONG) is an initiative which aims
to establish a ground-based network of automated 1m telescopes to obtain
high-precision radial-velocity measurements of nearby, bright stars and to study their
properties, primarily using the technique of asteroseismology. 
The targets being studied will be close enough that accurate parameters, such
as angular diameter, parallax, temperature, and abundances, are well known in
advance or can be easily determined.

The first SONG node consists of a 1m telescope produced by ASTELCO GmBH and is
installed at the Observatorio del Teide, Tenerife, Spain. There are two instruments: 
a lucky imaging photometer (\citealt{jesper}) and a high-resolution spectrograph 
(R $\sim$ 100,000)  installed at the Coud\'e focus in
a container next to the dome. The spectrograph, which is the instrument relevant for this paper, has been optimized for 
high-precision radial-velocity measurements and employs an iodine cell 
for wavelength reference. 
All operations are fully automatic and no operator is needed on site or 
remotely. For more details see \cite{song} \& \cite{mfa2}. The SONG radial-velocity pipeline is described in \cite{muher}.
In this paper we present the results of the first long test run with the SONG node on Tenerife.

\begin{table*}
\centering
\caption{Basic parameters of 46~LMi}
\begin{tabular}{lrc}
\hline\hline
Parameter & Measured value & Reference \\
\hline
$V$                       & 3.83                 & SIMBAD \cite{simbad} \\
Temperature \teff         & 4690$\pm50\,$K         & \cite{bubar} \\
Metal abundance \feh      & -0.1$\pm$0.1         & \cite{bubar} \\
Parallax (mas)            & $34.38\pm0.21$       & \cite{vanleeuwen} \\
Angular diameter (mas)    & $2.54\pm0.03$        & \cite{nordgren} \\
Radius $R$/\rsun            & 7.95$\pm$0.11        & from parallax and angular diameter\\
Surface gravity $\log g$  & $2.61\pm0.2$         & \cite{bubar} \\
Luminosity $L$/\lsun        & $27.42\pm1.38$         & from \teff\ and $R$ above \\
Rotation $v\sin(i)$       & $2.1$\,\kms & \cite{massarotti} \\
Possible age (see text)   & $2.7\pm0.5$\,Gy       & \cite{bubar} \\
\hline
\end{tabular}
\label{tab1}
\end{table*}

\section{The first SONG target 46~LMi \label{sect:46lmi}}

The bright ($V=3.83$) red giant 46~LMi (HR~4247, HD~94264, HIP~53229) with spectral type K0III was selected
as a suitable first target for SONG as the amplitudes and the timescale of the
solar-like oscillations in this star are large. These requirements were necessary to test the telescope and spectrograph performance.

The star also presents a good test case for asteroseismology as there are several 
high-precision measurements of its basic parameters  (Table\ \ref{tab1}).
This includes a well-determined parallax ($ \pi = 34.38\pm0.21$\,mas, \citealt{vanleeuwen}), 
and an angular diameter ($\Theta_L = 2.54\pm0.03$\,mas) measured by interferometry \citep{nordgren}.
Together, these lead to a classical measurement of the radius of $R = 7.95\pm0.11$\,\rsun.

In Table\ \ref{tab1} the selected values from the literature are listed. 
The estimate of the age comes from the possible membership of the moving 
group WOLF 630 \citep{bubar} of 46~LMi.

\section{Observations and data reduction\label{sect:obs}}

\begin{figure}
\centering
\includegraphics[width=9cm]{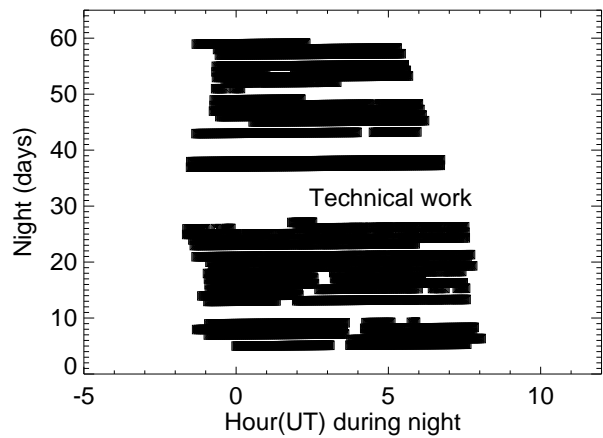}
\caption{Time coverage of the 46~LMi campaign. First night: 14 February 2014.
}
\label{techel}
\end{figure}

The SONG node permitted a very good time coverage, but with just one
node operational the data still suffer from gaps in the time series during
daytime. Nevertheless, we were able to obtain
a very long time series without involving any observer.

We  obtained a total of 3211 exposures, each with an exposure time of 240\,s and a 
CCD readout time of 2.5\,s. The
 observations were carried out from 14 February to 27 April 2014, with
 data obtained on 35 of the nights. The main causes for interruptions were
 poor weather conditions and technical work being done on the telescope and control system.
 All observations were carried out automatically; in the  late afternoon 
 we inserted the observing requests for the coming
 night into our queue system, and these requests were then executed during the night. All
 calibration files (bias, flats, ThAr reference) were obtained automatically 
 each afternoon prior to the start of the observations.
Figure\ \ref{techel} shows the distribution of data points over the
55-day period allocated to this target. The large gap is
due to a period of necessary technical upgrades; the
data were obtained while the telescope was being commissioned. The rest of the time 
the only interruptions were due to bad weather.

 The spectrograph has six possible slits, and for all observations we used the one
 providing a resolution of 90,000. For accurate wavelength calibration, we
 used an iodine cell. All spectra were bias corrected, flat-fielded and wavelength
 calibrated using the REDUCE package implemented in IDL (see \citealt{pis02})\footnote{{\tt http://www.astro.uu.se/}{$\sim $}{\tt pisku\-nov/RE\-SEARCH/REDUCE/}}.
 To extract  the precise radial velocities we used the  software iSONG (also IDL based)  (see e.g. \cite{cor12}, \cite{ant13}, and \cite{muher} for more details). There
 are a total of 24 useful orders with iodine lines available for calculating the velocities.

\begin{figure}
\centering
\includegraphics[width=8cm]{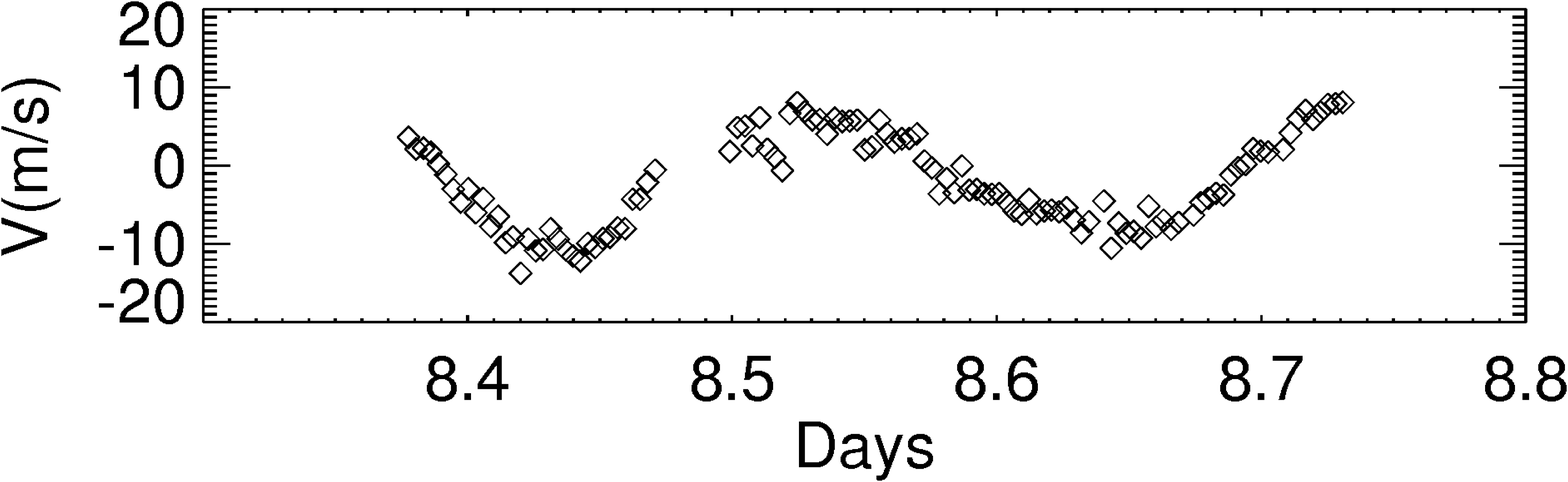}
\caption{Radial velocity curve of a single night.
}
\label{datastring}
\end{figure}

In  Fig.\ \ref{datastring} we present a full night of the radial-velocity
measurements. The typical characteristic timescale of variability of
the stochastic oscillations is clearly seen in the velocity curve as the noise level per data point is much lower
than the periodic signal.
Typically, the instrumental noise per data point is 1.5 ${\rm m\,s}^{-1}$ on a
good night, where the noise has been calculated as the scatter around a smooth curve
through the data. 

\section{Analysis of the radial-velocity time series \label{sect:vseries}}
\subsection{Frequency of maximum power:  $\nu_{\rm max}$}
\label{sec:numax}
\begin{figure*}
\centering
\includegraphics[width=16cm]{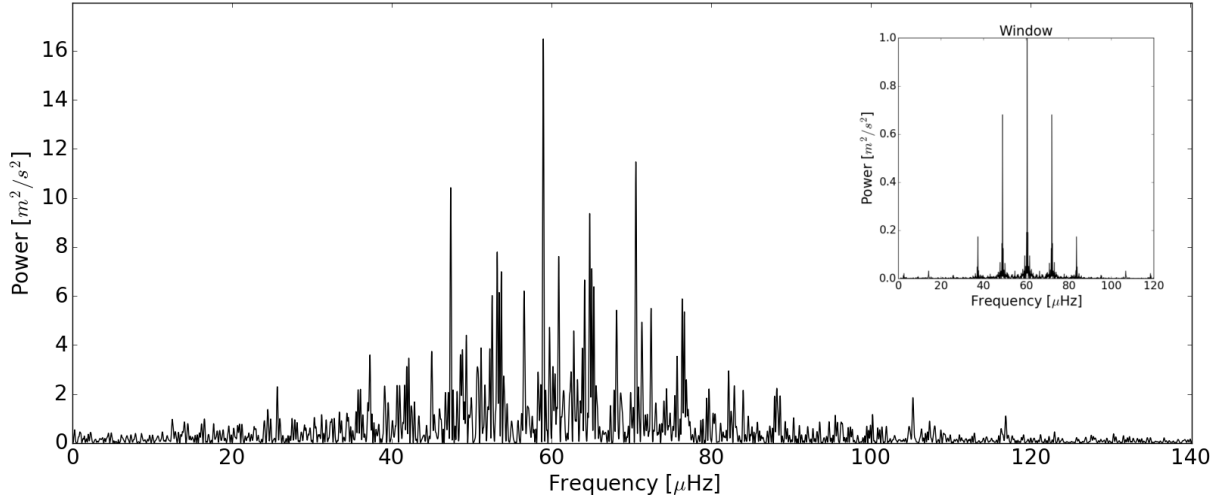}
\caption{High-pass filtered power spectrum of 46~LMi. The inset shows the normalized window function, which shows strong sidelobes.
}
\label{power}
\end{figure*}

The power spectrum (shown in Fig.\ \ref{power}) was calculated as a fit of sinusoids to the raw radial-velocity time series, as described in \cite{1995A&A...301..123F}.
The oscillation signal is clearly visible as the power excess around $60  \,\mu$Hz.
A low-frequency filtering has been applied where ten frequencies below $6  \,\mu$Hz were cleaned from the power spectrum.
The effect of single-site observations is illustrated by the window
function in the inset, which shows strong alias peaks at $\pm11.57 \,\mu$Hz corresponding 
to the daily gap in the time series. This significantly complicates the 
determination of the frequencies of the $p$ modes.

The frequency of maximum power ($\nu_{\rm max}$) is a widely used value which
relates to the large frequency separation through scaling relations \citep{stello2009, huber, Handberg2016, arentoft2017}.
Based on the available data we determined $\nu_{\rm max}$ from a Gaussian fit to the high-pass filtered power 
spectrum. The uncertainty on $\nu_{\rm max}$ was determined by splitting the
time series into seven chunks, each with the same number of data points. For each chunk the unfiltered power
spectrum was calculated and a value for $\nu_{\rm max}$ was determined by fitting a Gaussian
to the interval showing power excess, $30 \, \mu$Hz -- $100 \,\mu$Hz. The uncertainty on the 
final frequency of maximum power was then given as the standard deviation of the mean error.
This procedure yielded $\nu_{\rm max} = 59.4 \pm 1.4 \,\mu$Hz.
It is worth mentioning that the expected mode lifetime is of the order of the time span of the data set. This can have an effect on the determined $\nu_{\rm max}$, which will be highly influenced by the dominant mode in the power spectrum.

\begin{figure*}
\centering
\includegraphics[width=16cm]{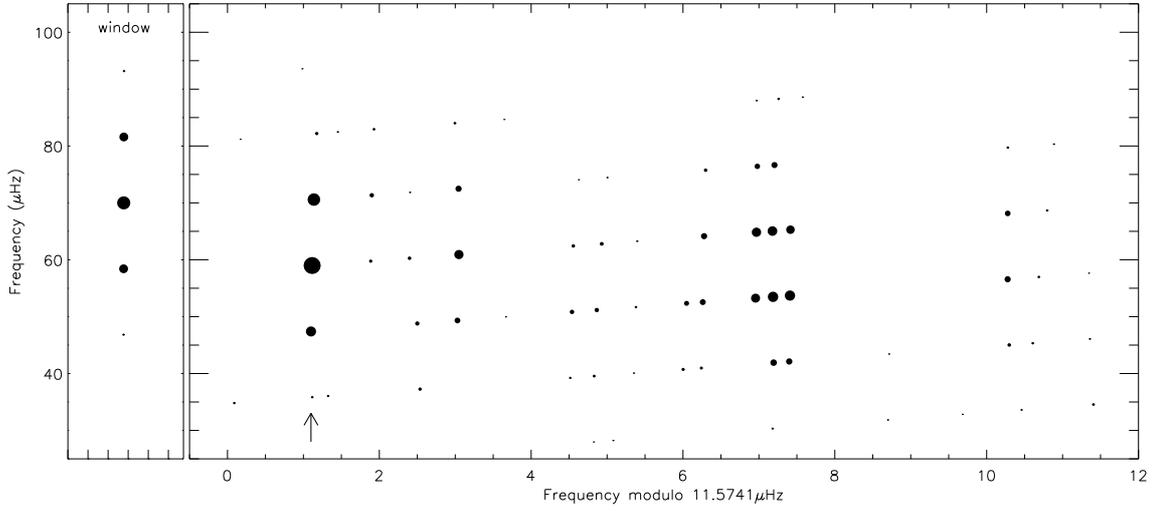}
\caption{\'Echelle diagram with respect to the daily alias $11.57 \,\mu$Hz, which
brings out peaks belonging to the window function. This is the basis
for locating a number of possible modes listed in Table~\ref{tabfrq}. The arrow points to the five peaks corresponding to the dominant mode in the power spectrum.
}
\label{echelle}
\end{figure*}

\subsection{Individual frequencies}

We have attempted to determine the frequencies of modes in the power spectrum
of 46~LMi using two techniques. 
In the first method, illustrated in Fig.~\ref{echelle},
peaks found in a smoothed power spectrum are plotted in an \'echelle diagram with frequency modulo $11.574\,\mu$Hz on the
abscissa and $\nu$ on the ordinate. 
Then modes are found by looking for patterns similar to the window function
in the vertical direction. The size of the symbols represents the
amplitude of each detected peak and can be used to select the
correct frequency, but in some cases  an alias might be obtained instead. Some of the peaks shown in the figure are statistically non-significant and are only included to look for the aliases in the window pattern.

In the second method  an iterative frequency determination is performed using PERIOD04. 
First, the frequency of the mode with maximum signal is found. A weighted cosine fit to the time series with the corresponding period is calculated and subtracted from the time series. This way side bands caused by the window function are removed in the power spectrum. This is repeated on the residual spectrum, but the following
iterations simultaneously fit multiple cosines to the time series. This operation is halted when  the same number of frequencies as in the first method has been reached. 
This method is possible because the resolution
in the power spectrum ($0.2\, \mu$Hz) is larger than the line width of the stochastic
modes in red giants ($\approx 0.1 \,\mu$Hz, \citealt{handberg2016b}). This means excited modes will be represented by only one peak in the power spectrum in most cases and not by a set producing a Lorentzian profile, which is the case when the observations span  many life cycles.\\
Finally, we get a list of modes (frequency, amplitude, and phase), 
where modes are only included if they are detected by both methods and agree
to better than 0.1~$\mu$Hz.
Both results ($\nu_1, \nu_2$) are presented in Table \ref{tabfrq} to give an idea
of the precision, and generally good agreement is seen. 
The amplitude $A_2$ is given for the PERIOD04 method. For
two frequencies PERIOD04 detects what could be an alias, which  we have enclosed in  parentheses. All modes included in Table~\ref{tabfrq} have high signal-to-noise ratios ($S/N$). More modes are likely present; however, for the remaining peaks we were not able to determine which are the true modes and which are the aliases, and they have therefore not been included.

\begin{table} \label{tab:modes}
\centering
\begin{threeparttable}[b]
\caption{Possible modes in 46~LMi}
\label{tabfrq}
\begin{tabular}{llll}
\hline\hline
 $\nu_1$ & $\nu_2$ &  $A_2$ \tnote{1} & S/N \tnote{2}\\
 ($\mu$Hz) & ($\mu$Hz) & (${\rm m\,s^{-1}}$) \\
\hline
      58.986  &    58.958   &   5.1 & 72.9\\
      53.705  &    53.703   &   2.3 & 32.9\\
      64.838  &    64.837   &   1.9 & 27.1\\
      60.919  &    60.914   &   1.9 & 27.1\\
      51.160  &    51.157   &   1.9 & 27.1\\
      64.148  &    64.143   &   1.7 & 24.3\\
      71.345  &    (82.904) &   --  & ${\sim}25$\\
      48.796  &    (60.451) &   --  & ${\sim}25$\\
\hline
\end{tabular}
\begin{tablenotes}
  \item[1] $\nu_1$ is the frequency from method 1. $\nu_2$ and $A_2$ the frequency and amplitude from method 2 (PERIOD4).
   \item[2] Noise level calculated at high frequency to 0.07m/s in PERIOD04.
\end{tablenotes}
\end{threeparttable}
\end{table}

\subsection{Solution using $\nu_{\rm max}$} \label{reality}

The global asteroseismic observables include $\nu_{\rm max}$ defined above and
the large frequency separation $\Delta \nu$, i.e.  the average frequency
spacing between modes of the same degree and adjacent orders.

As  presented by \cite{Miglio2012}, the asteroseismic scaling relations for mass can be expressed in the following four different ways if  the luminosity $L$ is included as an independent variable along with $T_{\rm eff}$, $\Delta\nu$, and $\nu_{\rm max}$:

$$\frac{M}{{\rm M}_\odot} \simeq \left({\frac{\Delta\nu}{\Delta\nu_\sun}}\right)^{-4}\left(\frac{\nu_{\rm max}}{\nu_{\rm max,\sun}}\right)^3\left(\frac{\Teff}{\Teffsun}\right)^{3/2} \; , $$ \\
$$\frac{M}{{\rm M}_\odot} \simeq \left(\frac{\Delta\nu}{\Delta\nu_\sun}\right)^{2}\left(\frac{L}{{\rm L}_\sun}\right)^{3/2}\left(\frac{\Teff}{\Teffsun}\right)^{-6} \; , $$ \\
$$\frac{M}{{\rm M}_\odot} \simeq \left(\frac{\nu_{\rm max}}{\nu_{\rm max,\sun}}\right)\left(\frac{L}{{\rm L}_\sun}\right)\left(\frac{\Teff}{\Teffsun}\right)^{-7/2} \; , $$ \\
$$\frac{M}{{\rm M}_\odot} \simeq \left(\frac{\nu_{\rm max}}{\nu_{\rm max,\sun}}\right)^{12/5}\left(\frac{\Delta\nu}{\Delta\nu_\sun}\right)^{-14/5}\left(\frac{L}{{\rm L}_\sun}\right)^{3/10} \; . $$ 
Using $L$ we exploit the asteroseismic scaling relation for mass in the form that does not involve $\Delta \nu$. We use the standard asteroseismic values and $T_{\rm eff}$ for the Sun \citep{Miglio2012}, which are $\Delta \nu_{\sun}=135\,\mu$Hz, $\nu_{\rm max,\sun}=3100\,\mu$Hz, and $T_{\rm eff,\sun}=5777\,K$. 

We adopted the luminosity ($L = 27.42 \pm 1.38$\,\lsun ) determined by combining the spectroscopic $T_{\rm eff}$ 
and radius, $R = 7.95\pm0.11$ \rsun\, from angular diameter and parallax distance since in that way we avoid making use of a bolometric correction. 
We  checked that the luminosity derived in an alternative way from the parallax $\pi=34.38$\,mas, the visual magnitude $V=3.83,$ and the bolometric corrections from \cite{casagrande2014}, $BC_{V} = -0.384$ and ${BC}_{V,\odot\ } = -0.068,$ are in agreement with the adopted values to well within the $1\sigma$ uncertainty.

With this approach we obtained $M = 1.09\pm0.04$\,\msun, after which reversing any of the other mass equations leads to $\Delta\nu = 6.30\pm0.08 \,\mu$Hz with the uncertainties
determined by bootstrapping \citep{feigelson2012modern}. \
With the estimate of the mass and radius we can determine the surface gravity of 46~LMi to $\log g = 2.674 \pm 0.013$.

It is well known  that a correction to the observed $\Delta\nu$ is needed before its use with the asteroseismic 
scaling relations \citep{Brogaard2016, Handberg2016, Miglio2012, mosser2013, guggenberger2016}. 
From figure 3 in \cite{rod2017} for a mass of 1.09 \msun\ and $\nu_{\rm max}=59.4\,\mu$Hz in the two cases with [Fe/H] = -0.25 and [Fe/H] = 0.00 to match the [Fe/H] = -0.1 of 46~LMi we get a correction factor of 0.966 as the mean of the two solutions. Applying this correction to $\Delta\nu = 6.30\pm0.08 \,\mu$Hz we obtain an estimate for the observed $\Delta\nu=6.09\pm0.09\,\mu$Hz. The increase in uncertainty comes from the metallicity dependence of the correction. 

\subsection{Estimating $\Delta\nu$ using other methods}

We tried to find the large frequency separation $\Delta\nu$ by plotting the detected modes in
an \'echelle diagram with different guesses for the large separation,
but no convincing $\Delta\nu$ could be found in this way due to the significant number of alias peaks. Another unsuccessful approach was to calculate the autocorrelation of the power spectrum, and we therefore tried the method that we now discuss here.

\subsubsection{Comparison to red giants observed by \kepler}
\label{sec:scaling}

\begin{figure}
\centering
\includegraphics[width=\columnwidth]{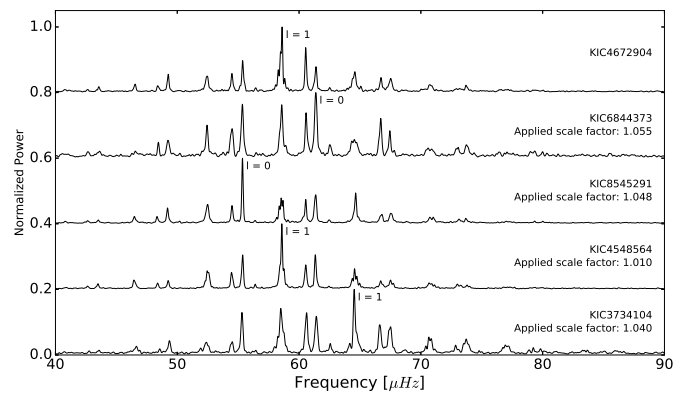}
\caption{Four red giants scaled to KIC 4672904. The scale factor applied to the frequency axis of each power spectrum is shown. The  l-mode label marks the mode identification of the dominant mode in the given spectrum.}
\label{fig:keplerscaled}
\end{figure}

We made use of the wealth of information available from the \kepler\ observations \citep{jenkins}.
From the APOKASC\footnote{APOKASC comes from a combination of Apache Point Observatory Galactic
Evolution Experiment (APOGEE) and of the Kepler Asteroseismic Science Consortium (KASC)} catalogue \citep{apokasc} we chose five red giants observed 
by \kepler\ with similar APOKASC effective temperatures and radii to 46~LMi (See Table~\ref{tabres}). 
The asteroseismic masses of the five stars are close to the value we determined in Sect.~\ref{reality} for 46~LMi.
The idea is to apply a scale factor to the power spectrum of a \kepler\ red giant to match the power spectrum of 46~LMi, but first
the method was tested on the \kepler\ stars alone.
The frequencies in the power spectrum of each \kepler\ star was multiplied by a variable scale factor, and the cross-correlation between the smoothed power spectrum and a chosen smoothed reference spectrum (KIC 4672904) was determined for each value of the scale factor. Each power spectrum was normalized with the power of the highest peak in the region of the oscillation signal. A first guess on the scale factor was based on the $\Delta\nu$ given in APOKASC and the cross-correlation was done for a range around this value. For each star the maximum in the cross-correlation function was found and this value of the scale factor was applied to the frequencies in the power spectrum. The result of this test (see Fig.~\ref{fig:keplerscaled})    demonstrates how well all modes align and proves that the method works for this sample of \kepler\ stars.
We also tested the method on a larger sample of more than 50 \kepler\ stars which were not as similar to 46~LMi, but we still  obtained a good agreement.\\
It is important to mention that applying the determined scale factor for the individual \kepler\ stars to the $\Delta\nu$ of the reference does not produce the exact same $\Delta\nu$ as given in APOKASC but they are well within the $1\sigma$ uncertainties. The scaled values are given in Table~\ref{tabres}.

\begin{figure}
\centering
\includegraphics[width=\columnwidth]{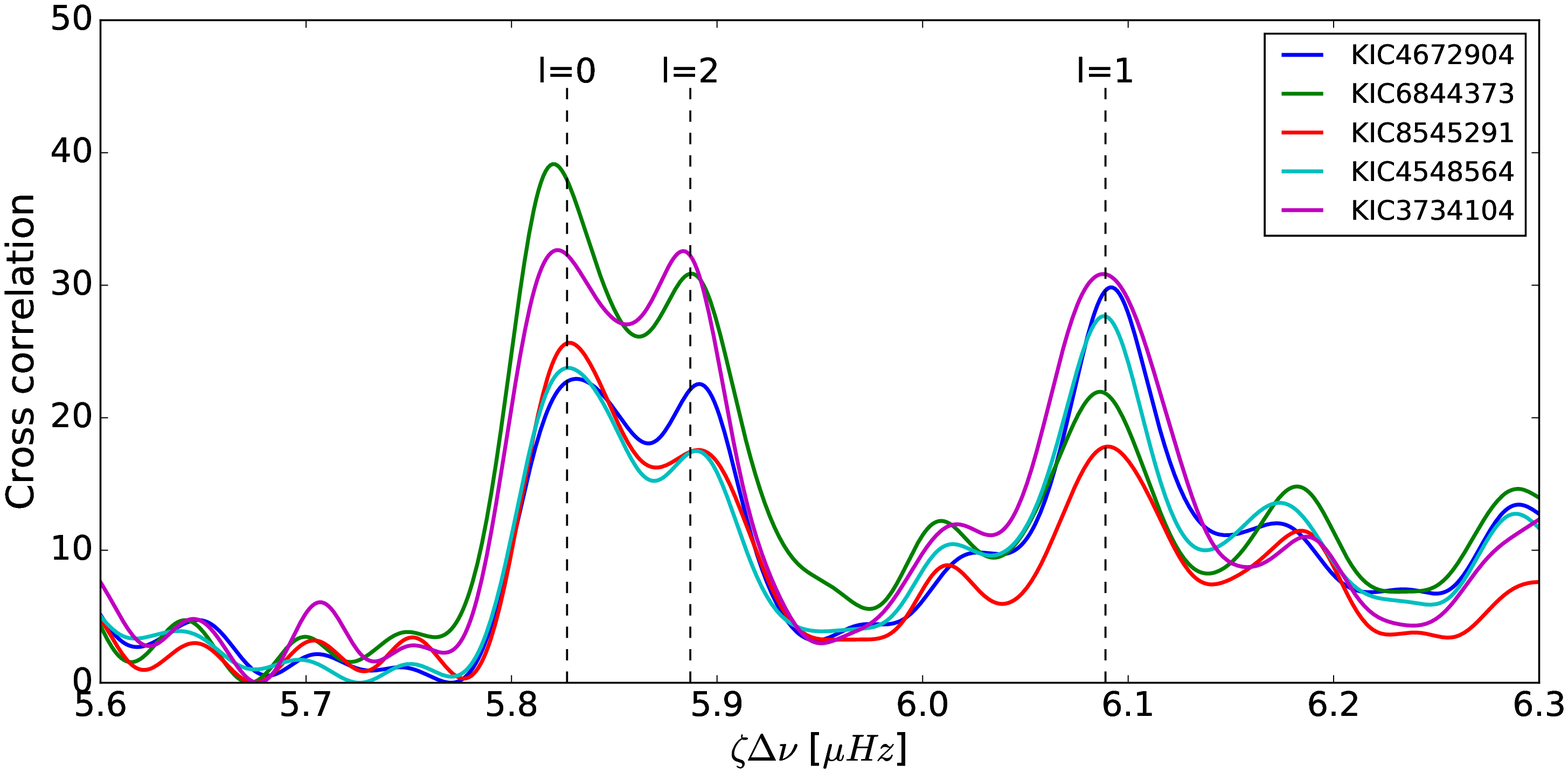}
\caption{Cross-correlation function for each of the \kepler\ stars correlated with 46~LMi. The abscissa is the scale factor multiplied by the $\Delta\nu$ of KIC 4672904.  The dashed lines indicate the different $l$-values coming from the identified modes in the \kepler\ stars when scaled to the dominant mode in 46~LMi. See the text for further information.}
\label{fig:scale-factor}
\end{figure}

\begin{figure}
\centering
\includegraphics[width=\columnwidth]{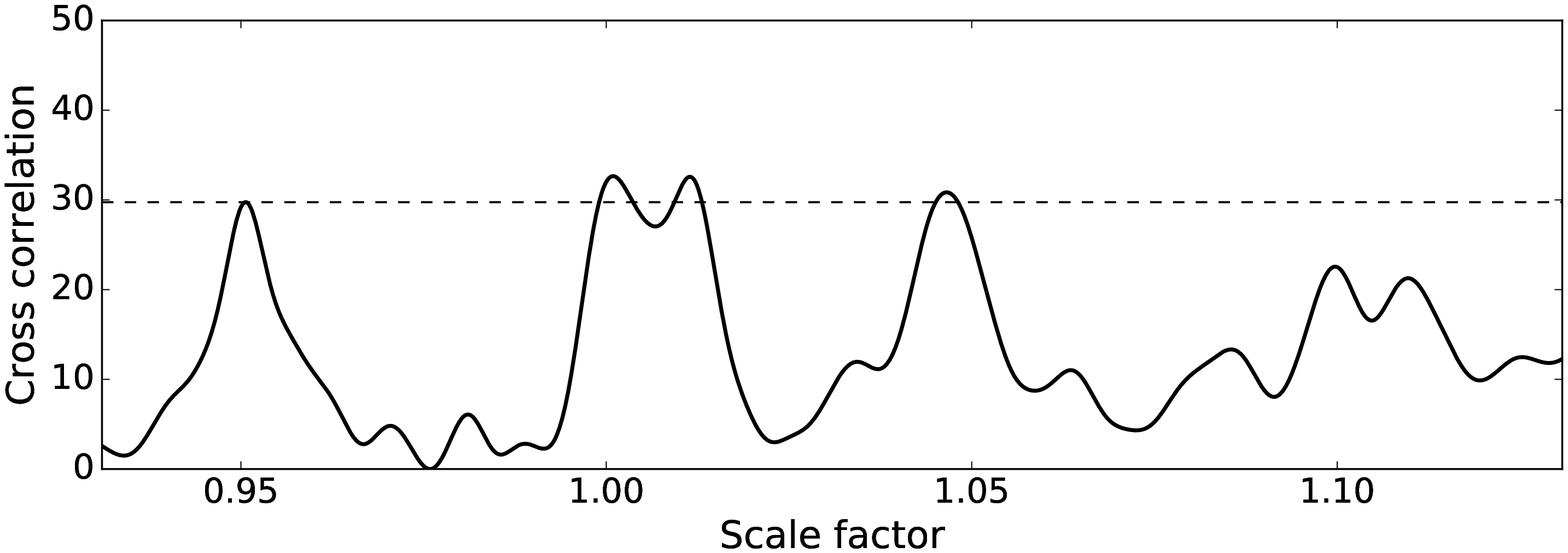}
\caption{Cross-correlation between KIC~3734104 and 46LMi for a larger range of scale factors. The peak at 0.95 corresponds to the match when the dominant mode in KIC~3734104 (see bottom panel in Fig.~\ref{fig:keplerscaled}) is scaled to the dominant mode in 46~LMi.}
\label{fig:KIC37-46LMi}
\end{figure}

The  method was then applied to 46~LMi which was used as the reference for the cross-correlation. In Fig.~\ref{fig:scale-factor} the cross-correlation function for each of the \kepler\ stars from Table~\ref{tabres} is shown. The correlation to 46~LMi is almost entirely dominated by the highest peak in the 46~LMi power spectrum producing a highly correlated signal when matched with a central peak (around 60 $\mu$Hz in Fig.~\ref{fig:keplerscaled}) in the \kepler\ red giants. This results in three dominant peaks in the cross-correlation corresponding to $l = 0, 1$, and 2 from the \kepler\ stars  scaled to the dominant peak in 46~LMi. In the case of KIC~8545291 and KIC~3734104 the dominant mode in the power spectrum is outside the central region (See Fig.~\ref{fig:keplerscaled}). Nevertheless, it is still one of the modes in the central region that produces the highest correlation. This is illustrated in Fig.~\ref{fig:KIC37-46LMi} where the cross-correlation between KIC~3734104 and 46~LMi is shown with a larger range of scale factors. The first peak in this figure corresponds to the match when the dominant mode in KIC~3734104 is scaled to the dominant mode in 46~LMi. Even though this mode is almost twice as high in the power spectrum the cross-correlation is still stronger around the central modes around 60 $\mu$Hz. 

The three possible values of $\Delta\nu$ from the correlation corresponding to the different l-values can be determined with high accuracy (better than $0.01\,\mu$Hz), but the real uncertainty on these values comes from the uncertainty in the $\Delta\nu$ from the APOKASC catalogue. We selected KIC 4672904 as our reference for the scaled $\Delta\nu$ in Table~\ref{tabres} because it is the one  closest to 46~LMi in the physical parameters. These values for $\Delta\nu$ are used to plot the correlation as a function of frequency on the abscissa on Fig.~\ref{fig:scale-factor}. This is simply done by multiplying the given $\Delta\nu$ with the scale-factor values found from the cross-correlations. This means that if we  selected another star as our reference, the abscissa could shift up to $0.05\,\mu$Hz (in the case of KIC 6844373).

From this scaling method we get three possible values for the large frequency separation for 46~LMi, which we take as the mean of the five scaled $\Delta\nu$ from Table~\ref{tabres} with the uncertainty given as the half of the full range in these values:\\

\noindent$\Delta\nu_{46LMi}=(5.85\pm0.03)\,\mu$Hz (Dominant mode in 46~LMi is $l$=0)

\noindent$\Delta\nu_{46LMi}=(5.92\pm0.03)\,\mu$Hz (Dominant mode in 46~LMi is $l$=2)

\noindent$\Delta\nu_{46LMi}=(6.12\pm0.03)\,\mu$Hz (Dominant mode in 46~LMi is $l$=1)\\

We cannot say which $\Delta\nu$ is the correct one because we do not know the identification of the dominant mode in 46~LMi, but by looking at each value we can argue which are more or less likely.
Remembering the value of $\nu_{\rm max}$ found in Sec.~\ref{reality}, and using this with the known physical parameters of 46~LMi and the asteroseismic scaling relations, we found $\Delta\nu=6.09\pm0.09\,\mu$Hz. From this, we can argue that the $\Delta\nu$ corresponding to the highest peak being a $l$=1 mode is the most likely. This solution has some implications which do not support it. When the power spectrum from any of the \kepler\ stars is scaled using the $l$=1 value, none of their radial modes near $\nu_{\rm max}$ match peaks in the power spectrum of 46~LMi. This seems unlikely since no \kepler\ red giant in our sample shows this behaviour. However, it is well known that radial modes have less power relative to the dipole modes in RV compared to intensity \citep{Bedding1996}, which again could favor this solution. The relative ratio between $l$=0 and $l$=1 modes for SONG was computed in \cite{Handberg2011} and was determined to be 1.35.
The solution where the dominant mode in 46~LMi is a $l$=0 mode is less likely \textbf{(${\sim}3\sigma$)} when compared to the $\Delta\nu$ estimated from $\nu_{\rm max}$ in Sec.~\ref{reality}. However, more modes from the power spectrum of a given \kepler\ giant  match peaks in the power spectrum of 46~LMi. This could be a result of the $\Delta\nu$ being very close to one half of the daily splitting produced by the sampling from single-site observations. Every second mode in the \kepler\ power spectrum would then roughly match an alias peak if this were the true value and the high correlation for the $l$=0 mode might simply be a result of the sampling of our data.\\
The same is the case for the value related to the dominant mode in 46~LMi being a $l$=2 mode. This value is also close to one half of the daily alias. The fact that it is a bit further away from the alias peaks might mirror the correlation power in Fig.~\ref{fig:scale-factor}, which is in general  a bit lower for the $l$=2 peaks than the $l$=0 modes.\\

The high peaks in Fig.~\ref{fig:scale-factor} have their origin in the dominant modes in the power spectra
of 46~LMi and the template. To reduce the effect of the dominant modes we have divided the power
spectra by a Gaussian fit to the power spectra before cross correlation. We have also performed a
test with a power law. In none of the cases did we see any significant change of the results.
The heights of the peaks in Fig.~\ref{fig:scale-factor} were modified, but not the positions.

All three values of $\Delta\nu$ can be used to scale a given \kepler\ star to 46~LMi, and all three fail to match all frequencies given in Table~\ref{tabfrq}. This indicates that we have either incorrectly determined some of the frequencies given or that none of the stars in our sample matches 46~LMi completely, or a combination of the two. We know from the \kepler\ red giants that some show clear splitting of the dipole modes caused by mixing with g-modes. If this were the case in 46~LMi it would make the analysis of the power spectrum significantly more difficult since all split modes would produce alias peaks due to the window function.

\begin{table*}
\centering
\caption{Large separation of 46 LMi, $\Delta\nu_{46 LMi}$ in $\mu$Hz from
scaling the large separation $\Delta\nu$ of \kepler\ giants}
\begin{threeparttable}
\begin{tabular}{lrrrrrrr}

\hline\hline
\kepler\ target & $\Delta\nu~(\mu$Hz)\tnote{a} & $\Delta\nu_{scaled}~(\mu$Hz)\tnote{b} & $\Delta\nu_{46Lmi}~(\mu$Hz)\tnote{c} & \teff (K)\tnote{a} & $M$/\msun\ \tnote{a} & $R$/\rsun\ \tnote{a} & \feh \tnote{a} \\
\hline
KIC 4672904 & $6.05\pm0.13$ & - & 5.82, 5.89, 6.09 & 4614 & $1.11\pm0.11$ & 8.19 & -0.38 \\
KIC 6844373 & $5.78\pm0.15$ & 5.73 & 5.87, 5.94, 6.14 & 4556 & $1.10\pm0.15$ & 8.41 & -0.05 \\
KIC 8545291 & $5.81\pm0.15$ & 5.78 & 5.86, 5.93, 6.13 & 4522 & $1.09\pm0.13$ & 8.34 & -0.13 \\
KIC 4548564 & $5.98\pm0.13$ & 5.99 & 5.84, 5.91, 6.11 & 4503 & $1.05\pm0.12$ & 8.09 & -0.33 \\
KIC 3734104 & $5.85\pm0.14$ & 5.82 & 5.86, 5.93, 6.13 & 4514 & $1.21\pm0.17$ & 8.60 & 0.02 \\
\hline
\end{tabular}
\begin{tablenotes}
  \item[a]{Values from \cite{apokasc}; \teff~\&~\feh~are the UNCORRECTED ASPCAP (G1)}
  \item[b]{Scaled from KIC 4672904}
  \item[c]{Scaled from the given \kepler\ star based on the $l=0$, 2, and 1 maximum in the correlation function, respectively.}
\end{tablenotes}
\end{threeparttable}
\label{tabres}
\end{table*}

\section{Modelling} \label{sect:model}

We  can show that a model exists that reproduces our new 
asteroseismic results and existing non-seismic measurements.
In addition,  we can find a `best' solution in terms of all constraints.

We employed the \textsc{ASTEC} evolution code \citep{jcd08a} for stellar 
evolution modelling computations, and the \textsc{ADIPLS} oscillation 
package \citep{jcd08b} for frequency calculations. \textsc{ASTEC} can 
evolve models up to the tip of the red giant branch, which is adequate 
for 46~LMi. The input physics of the ASTEC version that we used included the 
latest OPAL opacity tables \citep{igl96}, OPAL 2005 equation of 
state \citep{rn02}, and NACRE reaction rates \citep{ang99}. At low 
temperatures, opacities are obtained from \cite{fer05}. Convection is 
treated under the assumption of {mixing length theory} \citep{boh58}. 
We did not take rotation, diffusion, or convective overshoot into consideration 
in our calculation.

For model selection, we used a tool called DIAMONDS that performs Bayesian 
parameter estimation and model comparison by means of the nested sampling 
Monte Carlo (NSMC) algorithm \citep{cor14}. 
The mass, heavy-element abundance $Z$, and mixing-length parameter $\alpha$ are set 
as free parameters. 
The mass is restricted to the range of 0.95--1.25 \msun, $Z$ to the range 0.001 
-- 0.03, and $\alpha$ to  1 -- 3. 
Based on a Galactic chemical-evolution model 
\citep{carigi2000, pietr2004}
the hydrogen abundance $X$ is obtained as $X = -2.4 Z + 0.748$, which 
gives $X = 0.7$ when $Z = 0.02$ and a helium-to-metal enhancement 
ratio $\Delta Y / \Delta Z = 1.4$. \\
We use the observed parameters (with 
their uncertainties) $R$, \teff\ from Table~\ref{tab1}, and $\nu_{\rm max}$ 
from Sect.~\ref{sec:numax} as 
constraints to obtain the likelihood of a given model based on the $\chi^2$ 
value. The results are given in Table~\ref{modelres} where the errors are given as Bayesian errors which only
represent the capability of the model to fit the given observations.
We use a 68.3\% probability for the Bayesian credible intervals.

\begin{table}
\centering
\caption{Most likely model parameters given the constraints from observations.
See the text for explanation of the errors given.}
\begin{tabular}{lrr}
\hline\hline
 Parameter  &  Value   & $1\sigma$ error \\
\hline
$M$/\msun       &    1.09       &    0.03\\
$Z$             &   0.017       &   0.003\\
$\alpha$        &     2.07      &    0.11\\
$R$/\rsun       &     7.95      &    0.13\\
{\teff}\,(K)    &     4688      &      86\\
$L$/\lsun       &    27.3       &    2.2\\
Age\,(Gyr)      &     8.2       &    1.9\\
$\log g$                &     2.67  &   0.01\\
$\nu_{\rm max} \, (\muHz)$      &     59.4 &    2.0 \\
$\Delta\nu \, (\muHz)$ &   6.3 &    0.1\\
\hline
\end{tabular}
\label{modelres}
\end{table}

If we assume that we know $R$ and \teff, we can use the scaling relation for $\nu_{\rm max}$
to calculate $M$ which we did in Sec.~\ref{reality}. Not surprisingly, the best fitting model therefore has parameters consistent 
with the previous results in this paper. The $\chi^2$ of the best models is 1.36. The values given in Table 4 are from the Bayesian and they give  $\chi^2=0.87$.
Removing the constraint on $\nu_{\rm max}$ changes the results
very little.
The age determined here is significantly older than found by \cite{bubar}, but it is the expected age for a $\sim1.09$ \msun\ red giant. If 46~LMi were indeed a member of the moving group
Wolf 630, this would suggest a higher age for the group. More likely, 46~LMi is not a member.

\section{Conclusions} \label{sect:concl}

Even if SONG is not yet a network of telescopes, the use of just one dedicated
automatic telescope leads to ground-based results comparable to or better than previous asteroseismic campaigns. The reason is the easy way SONG can perform long time observations.
The main results are summarized here.
\begin{itemize}
\item{}
The first SONG observations demonstrate that we have perhaps obtained  the best ground-based red-giant power spectrum to date. 
46~LMi shows a clear solar-like power excess centred
at 59.4$\pm1.4 \,\mu$Hz, where the uncertainty is mainly due to the
stochastic excitation. The power reaches nearly zero  at low 
frequencies ($\leq 30 \,\mu$Hz) for the high-pass filtered radial-velocity time series. The power spectrum is affected by alias peaks due to the use of single-site data.
Comparison to stars observed by \kepler\ leads to three possible values for the large frequency separation.  
We therefore base our mass and radius estimates on $\nu_{\rm max}$ as the only asteroseismic measure and combine it with classical results, as described in Sect.~\ref{reality}.
\item{}
Our best estimate of the mass of 46~LMi (from the $\nu_{\rm max}$ solution) is $M=1.09\pm 0.04$\,{\msun} and the radius from classical estimates is $R=7.95\pm 0.11$\,{\rsun}.
The surface gravity estimate is $\log g = 2.674 \pm 0.013$ derived from the mass and radius just listed leading to an improved accuracy
compared to the spectroscopic gravity $\log g = 2.61\pm 0.2$.
\item{}
The age of $8.2\pm1.9$\,Gyr that we find for 46~LMi is significantly higher than that reported by \cite{bubar}. 
This suggests that 46~LMi is not a member of the moving group WOLF 630.
We thus demonstrate that asteroseismic age determination can be used to determine membership of stellar constellations such as moving groups and clusters.
\end{itemize}

Improvements will be possible from multisite observations, as we will be able to identify individual mode frequencies since alias peaks will be highly suppressed. 
This will lead to a more precise age estimate, and a measurement of mass and 
radius from the asteroseismic scaling relations that is independent of the distance. 
We will then be able to get a high-accuracy test of the asteroseismic scaling relations by comparing the radius derived from asteroseismology to that obtained from interferometry. This clearly shows why more SONG nodes are needed.

\begin{acknowledgements}
This research took advantage of the {\it Simbad} and {\it Vizier} databases, operated  at the CDS, Strasbourg (France), and NASA's Astrophysics Data System Bibliographic Services. We would like to acknowledge the Villum Foundation, The Danish Council for Independent Research | Natural Science and the Carlsberg Foundation for the support on
building the SONG prototype on Tenerife.
K.B. acknowledges  support from the Villum Foundation.
The Stellar Astrophysics Centre is funded by The Danish National Research Foundation (Grant DNRF106) and research was supported by the ASTERISK project (ASTERo-seismic Investigations with SONG and Kepler) funded by the European Research Council (Grant agreement n. 267864). We also gratefully acknowledge the support from the Spanish Ministry of Economy Competitiveness (MINECO) grant AYA2016-76378-P.
\end{acknowledgements}

\bibliographystyle{aa}
\bibliography{lmi46-v19} 




\end{document}